\documentclass{article}
\usepackage[latin9]{inputenc}
\usepackage{booktabs}
\usepackage{textcomp}
\usepackage{amsmath}
\usepackage{amssymb}
\usepackage{graphicx}
\usepackage{esint}
\usepackage[numbers]{natbib}

\makeatletter

\newcommand{\lyxmathsym}[1]{\ifmmode\begingroup\def\b@ld{bold}
  \text{\ifx\math@version\b@ld\bfseries\fi#1}\endgroup\else#1\fi}

\providecommand{\tabularnewline}{\\}

\usepackage{amsfonts}
\usepackage{cite}

\newcommand{\comment}[1]{}

\date{}

\makeatother

\begin{document}
\comment{
\begin{quote}
Title: \emph{Memory controls time perception and intertemporal choices}

\bigskip

\emph{Pedro A. Ortega}\\
School of Engineering and Applied Sciences\\
University of Pennsylvania, Philadelphia, U.S.A. \\
Email: pedro.ortega@gmail.com

\bigskip

\emph{Naftali Tishby} (Corresponding Author) \\
The Selim School of Computer Science and Engineering \\
\& The Edmond and Lilly Safra Center for Brain Sciences\\
The Hebrew University, Jerusalem, Israel. \\
Email: tishby@cs.huji.ac.il \newpage{}
\end{quote}
} % End of comment.

\title{Memory shapes time perception and intertemporal choices}

\author{%
\begin{tabular}{cc}
Pedro A. Ortega & Naftali Tishby\tabularnewline
University of Pennsylvania & The Hebrew University in Jerusalem\tabularnewline
ope@seas.upenn.edu & tishby@cs.huji.ac.il\tabularnewline
\end{tabular}}
\maketitle
\begin{quote}
There is a consensus that human and non-human subjects experience
temporal distortions in many stages of their perceptual and decision-making
systems. Similarly, intertemporal choice research has shown that decision-makers
undervalue future outcomes relative to immediate ones. Here we combine
techniques from information theory and artificial intelligence to
show how both temporal distortions and intertemporal choice preferences
can be explained as a consequence of the coding efficiency of sensorimotor
representation. In particular, the model implies that interactions
that constrain future behavior are perceived as being both longer
in duration and more valuable. Furthermore, using simulations of artificial
agents, we investigate how memory constraints enforce a renormalization
of the perceived timescales. Our results show that qualitatively different
discount functions, such as exponential and hyperbolic discounting,
arise as a consequence of an agent's probabilistic model of the world.


\emph{Significance Statement:} We propose that perceived durations
can be quantified in terms of an agent's memory changes due to the
encoding of past-future sensorimotor dependencies. Consistent with
findings in psychophysics, we show that events that are predicted
to be unlikely yet rewarding are perceived as being longer in duration.
Furthermore, through the simulation of artificial agents, we show
that the asymptotic behavior of intertemporal preferences can be explained
as a consequence of the model class employed by an agent to predict
its future.
\end{quote}

\section{Introduction}

Our aim is to propose a model of subjective time based on information
theory and to investigate its implications relative to two phenomena:

\paragraph{Time perception. }

Why does time appear to slow down when you visit a new place, and
speed up once you get familiar with it? Recent findings in psychology,
neuroscience, and ethology suggest that perceived duration does not
coincide with physical duration, but rather depend on the statistical
properties of stimuli. Experiments in psychophysics experiments have
shown that, if presented with a train of repeated stimuli at constant
time intervals (e.g., a letter, word, object, or face), subjects would
perceive them as decreasing in duration \citep{Tse2004,Pariyadath2008}.
On the other hand, the opposite effect is reported whenever the properties
of a train of stimuli are suddenly changed: brighter \citep{Brigner1986,Terao2008},
bigger \citep{Ono2007,Xuan2007}, dynamic \citep{Brown1931,Kanai2006},
or more complex stimuli \citep{Schiffman1974,Roelofs1951} appear
to last longer. Measurements of brain activity have found that longer
durations correlate with increased neuronal firing rates, fMRI, or
EEG signals \citep{Barlow1978,Dupont1994,DeJong1994,Linden1999,Ranganath2003,Murray2006}.
When combined with ideas from information theory, these observations
have led to the hypothesis that the subjective duration of a stimulus
is proportional to the amount of neural energy required to represent
said stimulus, and that this energy is a signature of the coding efficiency
\citep{Eagleman2009}.

\paragraph{Intertemporal choice.\emph{ }}

Why do \$100 today feel more than \$100 tomorrow? Intertemporal choices
lie at the heart of economic decision-making. The economic literature
has proposed early on \citep{Samuelson1937} that rational decision-makers
prefer immediate rewards over similar rewards in the future because
they \emph{discount time}---that is, future moments in time weight
less in their assessment of utility. Historically, the first and most
commonly used mathematical model of temporal discounting, namely \emph{exponential
discounting}%
\footnote{Also known as \emph{geometric discounting} in the artificial intelligence
literature.%
}, proposes that decision-makers discount future utilities using a
single, constant discount rate which compresses distinct psychological
motives \citep{Samuelson1937,Frederick2002}. The second most influential
type of model is \emph{hyperbolic discounting}, which postulates that
discount rates \emph{decline} over longer time horizons, i.e.~future
values decrease less rapidly than exponential. Structurally, hyperbolic
discounting does not possess many of the elegant properties of exponential
discounting (such as e.g.~dynamical consistency); however, it has
significantly more empirical support \citep{Bradford2014,Laibson1997,Berns2007,Soman2005}. 

Temporal discounting was originally conceived as a property of the
decision-maker's preferences (e.g.~the utility function's curvature).
However, relatively recent studies have investigated the role of subjective
time as the underlying cause of intertemporal preferences. For instance,
prior work proposed that hyperbolic discounting arises due to a sub-additive
perception of duration \citep{Read2001}; and recent experimental
findings \citep{Takahashi2005,Zauberman2009,Bradford2014} have suggested
that intertemporal choice patterns are well captured by treating time
as a perceptual modality subject to classical psychophysical laws
(e.g. Weber-Fechner law). \bigskip{}

Similar to prior proposals \citep{Staddon2005}, the question we address
here is: how does an agent's memory affect time perception and intertemporal
preferences? Here we propose a model of time perception in terms of
an agent's memory requirements to encode (temporal) sensorimotor dependencies.
This approach does not touch upon the cognitive processes that implement
the sense of time (such as \emph{internal clock} and \emph{attentional
counter} models \citep{Gibbon1984,Matell2000,Dragoi2003,Maniadakis2014}),
nor does it attempt to provide a phenomenological account \citep{Suddendorf1997,Wheeler1997,DArgembeau2005,DroitVolet2009}.
Instead, by restricting our attention to the purely statistical properties
of behavior, we obtain a model of time perception that is agnostic
to the implementation details and the substrate. In this abstraction,
a range of tools become available that enable the quantitative investigation
of representational limitations. Our main finding is that a system's
memory simultaneously shapes its perception of time and its intertemporal
preferences, consistent with previous experimental findings.

\begin{figure}
\begin{centering}
\includegraphics[width=0.9\textwidth]{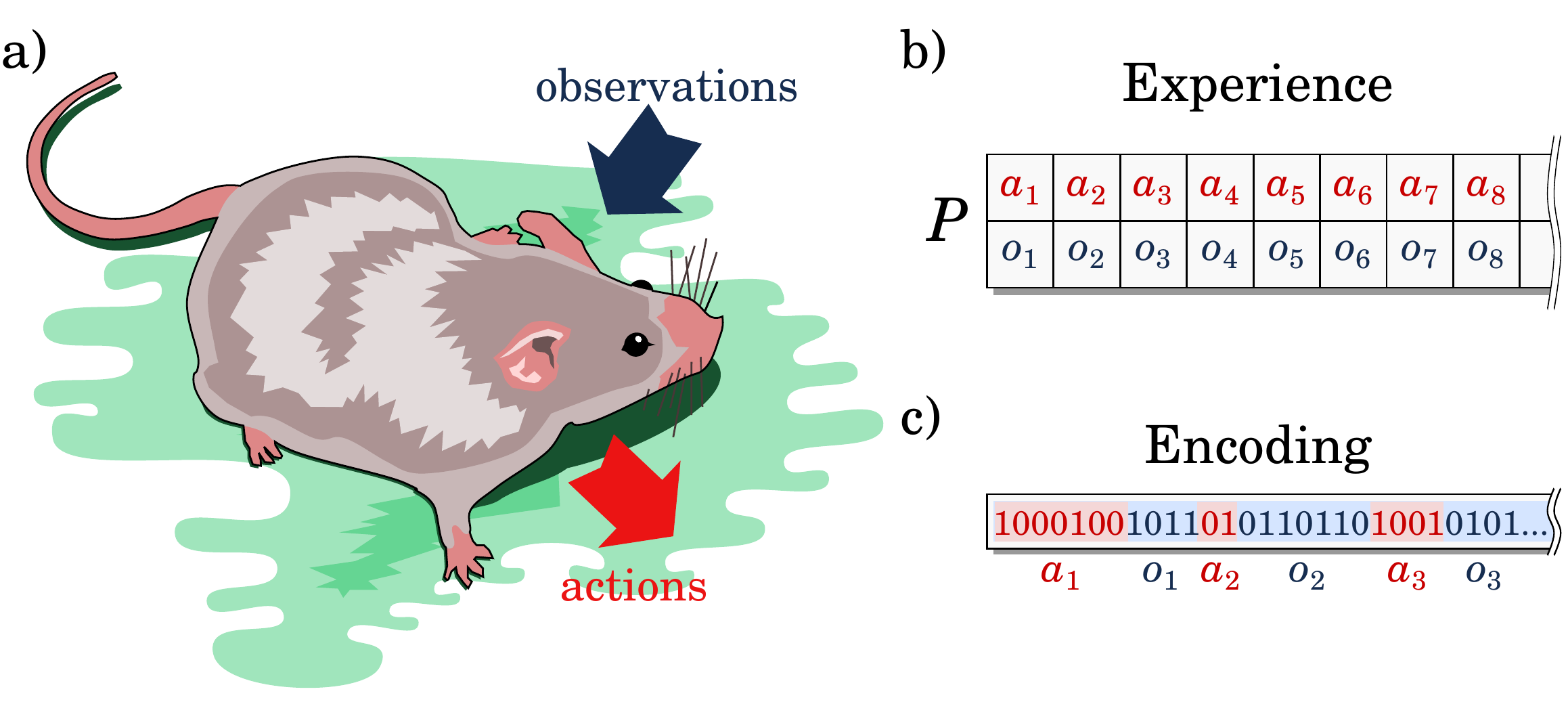}
\par\end{centering}

\caption{The interactions between an agent and its environment can be analyzed
using information theory \citep{Ashby1956}. (b) If we discretize
time and the continuum of sensorimotor events, then the interactions
can be thought of as a sequence of symbols being written on a tape,
generated by a stochastic process $P$. The symbols $a_{t}$ and $o_{t}$
are the action and the observation in turn $t$ respectively, and
$s_{t}$ is the corresponding memory state of the process. (c) As
is typical in information theory, we translate these symbols into
binary codewords to emphasize the complexity of generating them spontaneously
using fair coin flips. Specifically, a codeword of length $l$ corresponds
to a interaction occurring with probability $2^{-l}$ (panel~c).
The advantage of this view is that it enables a quantitative analysis
of the limitations on the process, such as the constraints on the
dependencies linking interactions at different points in time. \label{fig:organism-b}}
\end{figure}

\section{Predictive Information and Free Energy}

We consider adaptive agents that interact with an environment, e.g.
a mouse chasing food in a laboratory maze, a vacuum-cleaning robot,
or a ribosomal complex synthesizing proteins in the citoplasm (see
Fig.~\ref{fig:organism-b}). In artificial intelligence, such agent-environment
systems are often approximated using discrete-time stochastic processes
in which the agent and the environment take turns to exchange actions
and observations drawn from appropriately defined finite sets $\mathcal{A}$
and $\mathcal{O}$ respectively~\citep{RussellNorvig2009}. In each
time step, the agent acts following a policy $P(a|s)$ which specifies
the probability of generating the action $a\in\mathcal{A}$ when it
is in state $s\in\mathcal{S}$. Similarly, the environment replies
with an observation $o\in\mathcal{O}$ following the stochastic dynamics
$P(o|s,a)$. The agent generates actions so as to bring about its
goals, such as optimizing a feedback signal or maintaining a homeostatic
equilibrium. 

We center our analysis on the stochastic process describing the interaction
dynamics as seen from the perspective of the agent. An adaptive agent
has to simultaneously solve two learning tasks: predicting the environment
and learning to act optimally. In Bayesian reinforcement learning
\citep{Duff2002}, the standard approach is to model this as an agent
that uses past data to learn a parameter $\theta\in\Theta$ that encapsulates
both the properties of the environment and its corresponding optimal
policy. As the agent gains knowledge about $\theta$, it improves
its predictions about the environment and simultaneously its ability
to choose better actions (see Section~\ref{sec:results} for concrete
examples). Accordingly, the stochastic process $P$ represents the
agent's beliefs about the interaction dynamics, and the state~$s\in\mathcal{S}$
is given by the agent's \emph{information state} or \emph{memory state}%
\footnote{Notice that the agent's memory is not localized within its body: rather,
it is distributed between the body and the surroundings. For instance,
a vacuum-cleaning robot with a handful internal states can use the
objects in its environment as an external memory device---opening
the possibility to Turing-complete behavior. Hence, a comprehensive
analysis of the agent's behavior must be based on the memory of the
combined agent-environment system.%
}. We will assume that the state $s$ it is the minimal sufficient
statistic, i.e.~the minimal information shared between the past experience
and the parameter $\theta$, as any redundant information contained
in the past does not improve the agent's prediction abilities and
thus can be discarded.

In order to establish a link between the agent's memory state, reward
function, and temporal perception properties, we first need to clarify
what we mean by memory and reward, and how to infer these from the
stochastic process $P$.

\subsection{Predictive Information}

We first focus on the memory of the stochastic process. As is customary
in information theory, the information content of a given finite sequence
is assessed in terms of its binary codeword length~\citep{Cover1991}.
The binary length is a standardized proxy for the complexity of a
sequence, as it characterizes both the amount of two-state storage
units required for remembering it and the difficulty of generating
it from fair coin flips~\citep{Mackay2003} (Fig.~\ref{fig:organism-b}b).
Shannon proved that the optimal expected codeword length is given
by the entropy, implying that a minimal codeword has length $-\log p$,
where~$p$ is the probability of the sequence~\citep{Shannon1948}.
Many techniques exist to construct near-optimal lossless codewords
for data streams; a particularly elegant one is \emph{arithmetic coding~}\citep{Rissanen1976,Steinruecken2014}. 

What is memory? Following \citet{Bialek2001b}, imagine that we have
measured the first few interactions, and call this past~$x_{\text{p}}$.
Given $x_{\text{p}}$, we want to predict a finite number of future
interactions $x_{\text{f}}$. Even before looking at the data, we
know that some futures are more likely than others, and this knowledge
is summarized by a prior distribution $P(x_{\text{f}})$. If in addition
we take into account the information contained in the past, then we
obtain a more tightly concentrated posterior distribution over futures,
$P(x_{\text{f}}|x_{\text{p}})$. The number of bits in the past $x_{\text{p}}$
that are used to improve the prediction of a future $x_{\text{f}}$
is quantified as the difference between the prior and posterior codeword
lengths:
\begin{equation}
-\log P(x_{\text{f}})-\bigl(-\log P(x_{\text{f}}|x_{\text{p}})\bigr)=\log\frac{P(x_{\text{f}}|x_{\text{p}})}{P(x_{\text{f}})}.\label{eq:mi-pointwise}
\end{equation}
\ref{eq:mi-pointwise} measures the minimal memory that a system must
possess to enable this prediction%
\footnote{This difference can be negative. In this case, it can be interpreted
as the amount of information contained in the past $x_{\text{p}}$
that is inconsistent with the future $x_{\text{f}}$. %
}. If we average over all realizations, we obtain the mutual information
between the past and the future: 
\begin{equation}
I(X_{\text{p}};X_{\text{f}})=E_{P}\left[\log\frac{P(X_{\text{f}}|X_{\text{p}})}{P(X_{\text{p}})}\right].
\end{equation}
In the literature, this quantity is known as the \emph{predictive
information} \citep{Bialek2001a,Bialek2001b}. Intuitively, the predictive
information can be thought of as quantifying the amount of memory
that is preserved by the patterns, rules, or correlations that relate
the past with the future. An important property of the predictive
information is that it is subextensive if the stochastic process is
stationary. In other words, the predictive information has a sublinear
asymptotic growth in the length of a realization. This is in stark
contrast to the entropy of the process, which grows linearly with
the length of a realization. Consequently, only a vanishing fraction
of the dynamics of the process is governed by patterns; most of it
is driven by pure noise.

\subsection{Free Energy\label{sub:fe}}

Our second step is to establish a firm link between the statistics
of stochastic processes and their implicit rewards. The evolution
of the stochastic process can be analyzed using decision-theoretic
tools by adopting a complementary view to the previous one. In this
interpretation, conditioning amounts to imposing constraints on an
otherwise free evolution of the stochastic process. The assumption
is that, if uncontrolled, the future interactions $x_{\text{f}}$
would follow the stochastic dynamics described by the prior distribution
$P(x_{\text{f}})$. However, when the agent experiences the past interactions
$x_{\text{p}}$, it acquires knowledge that leads it to steer the
process into a more \emph{desired} direction, resulting in the dynamics
given by the posterior distribution~$P(x_{\text{f}}|x_{\text{p}})$~\citep{Todorov2009}.
This transformation can be characterized as the result of maximizing
expected rewards subject to constraints on the memory capacity of
the process \citep{Tishby2011,Ortega2013}. The associated objective
function is the free energy functional

\begin{align}
F(x_{\text{p}})[\tilde{P}]: & =\sum_{x_{\text{f}}}\tilde{P}(x_{\text{f}}|x_{\text{p}})\Bigl[R(x_{\text{f}}|x_{\text{p}})+F(x_{\text{p}},x_{\text{f}})\Bigr] &  & \text{(Expected Rewards)}\nonumber \\
 & \phantom{{=}}-\frac{1}{\beta}\sum_{x_{\text{f}}}\tilde{P}(x_{\text{f}}|x_{\text{p}})\log\frac{\tilde{P}(x_{\text{f}}|x_{\text{p}})}{P(x_{\text{f}})} &  & \text{(KL-Divergence})\label{eq:free-energy}
\end{align}
which is to be maximized w.r.t.~to the distribution $\tilde{P}$
over futures $x_{\text{f}}$ conditioned on the past $x_{\text{p}}$.
In the first expectation, $R(x_{\text{f}}|x_{\text{p}})$ is the \emph{real
reward}%
\footnote{We assume that rewards are additive:~$R(v,w|u)=R(v|u)+R(w|u,v)$.%
} of the future sequence $x_{\text{f}}$ conditioned on the past $x_{\text{p}}$
and $F(x_{\text{p}},x_{\text{f}})$ is the terminal reward of the
sequence $x_{\text{p}}x_{\text{f}}$. The second term is a penalization
that measures the memory cost of changing the probability of $x_{\text{f}}$.
The parameter $\beta>0$ is the \emph{inverse temperature,} and it
encapsulates the trade-off between rewards and information costs:
larger values correspond to cheaper memory costs and therefore more
control. The posterior distribution over the future is then defined
as the maximizer of~\eqref{eq:free-energy} given by the \emph{Gibbs
distribution}
\begin{equation}
P(x_{\text{f}}|x_{\text{p}}):=\frac{1}{Z}P(x_{\text{f}})\exp\Bigl\{\beta[R(x_{\text{f}}|x_{\text{p}})+F(x_{\text{p}},x_{\text{f}})]\Bigr\}\label{eq:gibbs-distribution}
\end{equation}
where $Z$ is a normalizing constant. 

Since \eqref{eq:free-energy} and \eqref{eq:gibbs-distribution} must
hold for any past-future window, two consequences follow. The first
is that the free energy functional has a recursive structure given
by the equality between optimal free energies%
\footnote{From an economic point of view, the optimal free energy \eqref{eq:ce}
turns out to be the \emph{certainty-equivalent value} of knowing~$x_{\text{p}}$;
in other words, it is the \emph{net worth} the agent attributes to
the future when it has experienced the past $x_{\text{p}}$.%
} and terminal rewards: 
\begin{equation}
F(x_{\text{p}})=\max_{\tilde{P}}F(x_{\text{p}})[\tilde{P}].\label{eq:ce}
\end{equation}
In particular, the terminal rewards $F(x_{p},x_{\text{f}})$ that
appear in equation \eqref{eq:free-energy} are themselves the result
of optimizing free energy functionals over the distant futures that
occur after $x_{\text{p}}x_{\text{f}}$. If we average over the pasts,
we get
\begin{equation}
\mathbb{E}\bigl[F(X_{\text{p}})\bigr]=\mathbb{E}\bigl[R(X_{\text{f}}|X_{\text{p}})+F(X_{\text{p}},X_{\text{f}})\bigr]-\frac{1}{\beta}I(X_{\text{p}};X_{\text{f}}),\label{eq:exp-fe}
\end{equation}
which reveals that the KL-penalization term is a constraint on the
predictive information, imposing a limit on the memory of the stochastic
process during the maximization of the expected rewards. The second
consequence is that rewards and log-likelihoods are related via an
affine transformation

\begin{equation}
\log P(x_{\text{p}}|x_{\text{f}})=\beta\Bigl[R(x_{\text{f}}|x_{\text{p}})+F(x_{\text{p}},x_{\text{f}})\Bigr]+C,\label{eq:likelihood-rewards}
\end{equation}
where $C$ is a constant. Intuitively, this means that ``the future
$x_{\text{f}}$ has more reward given the past $x_{\text{p}}$'' and
``the future $x_{\text{f}}$ is more likely given the past $x_{\text{p}}$''
are two equivalent statements%
\footnote{Similar points regarding the equivalence between rewards and likelihoods
were made previously, see e.g.~\citep{Friston2012a,Schwartenbeck2015}.
It is also worth clarifying that we treat rewards as \emph{internal,
subjective} quantities. This is consistent with expected utility theory
\citep{Neumann1944,Savage1954}, but unlike the more recent interpretation
in reinforcement learning where rewards are treated as externally
supplied, objective quantities. %
}. In addition, \eqref{eq:likelihood-rewards} provides a simple formula
for estimating the rewards from the realizations of a given stochastic
process.

\section{Results\label{sec:results}}

We have conducted simulations of agent-environment systems to achieve
two goals: first, to measure the agents' subjective time frame (made
precise later in this section); and second, to calculate their implicit
discount functions. The simulation results (samples from the stochastic
processes) provided us with the necessary data to subsequently estimate
(using the tools reviewed in the previous section) the implicit memory
constraints and rewards contained in the agent-environment interactions. 

For simplicity, our simulations are based on the standard framework
of multi-armed bandit problems \citep{Lai1985}. In these problems,
an agent gambles a slot-machine with multiple arms. When played, an
arm provides the agent with a Bernoulli-distributed nominal reward,
where the bias is initially unknown to the agent. The objective of
the game is to play a sequence of arms in order to maximize the sum
of rewards. Although simple, bandit problems pose many of the core
challenges of sequential decision-making. For instance, an agent has
to balance greedy choices versus choices intended to acquire new knowledge---a
trade-off known as the exploration-exploitation dilemma \citep{Sutton1998}.

\begin{figure}
\begin{centering}
\includegraphics[width=1\textwidth]{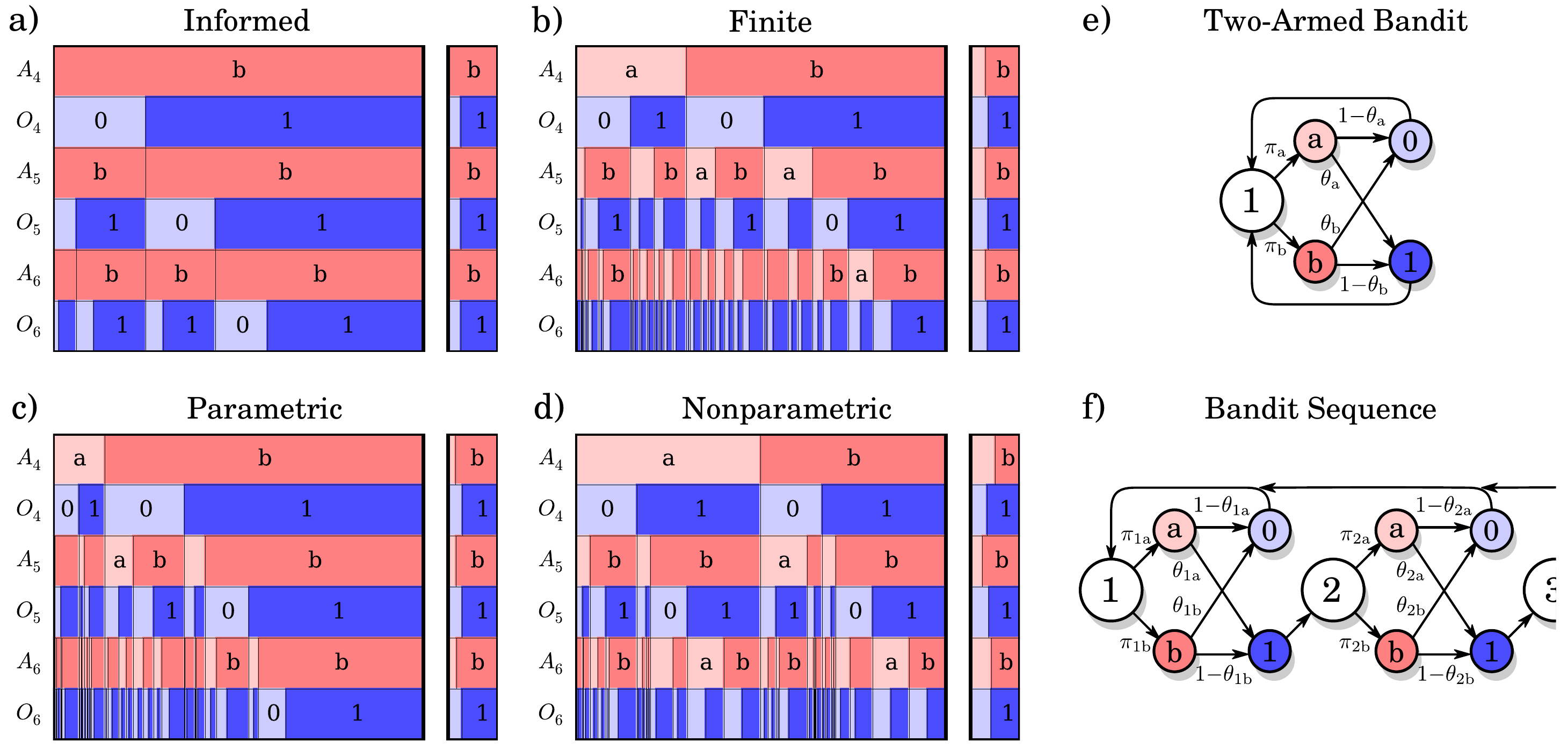}
\par\end{centering}

\caption{Predicted futures for agents with increasing model complexity. Agents
play a two-armed bandit with Bernoulli-distributed rewards. (a--d)
show their predictions for the next three interactions after having
experienced four initial interactions $x_{\text{p}}=(\text{a1,a0,b1,b1})$,
i.e.~where the agent switched to action ``b'' after observing a loss
in the second trial of arm ``a''. The diagrams show the predicted
probability (cell width) of each path of interactions, and sidebars
contain the marginal probabilities. In spite of playing the same bandit
depicted in (e), their predictions vary due to their probabilistic
models, ranging from an informed (a) to a nonparametric model (d).
The informed agent knows the Bernoulli biases of each arm and therefore
always makes perfect predictions and plays optimal actions. The finite
and parametric agents learn the transition probabilities of a single
two-armed bandit (e), whereas the nonparametric agent attempts to
fit a more complex model that is based on a sequence of bandits (f).\label{fig:model_classes}}
\end{figure}

Throughout all our simulations we used two-armed bandits with arms
labeled as ``a'' and ``b'' (Fig.~\ref{fig:model_classes}e). To investigate
how the learning ability impacts time perception, we simulated agents
with probabilistic models of increasing complexity (Materials \& Methods).
To do so, we used four different types of parameter spaces $\Theta$:
a singleton set~$\Theta$, representing an informed agent that already
knows the dynamics of the environment and the optimal policy; a finite
set~$\Theta$; a finite-dimensional parameter space $\Theta$ giving
rise to a parametric probabilistic model; and an infinite-dimensional
parameter space $\Theta$. We call these agents \emph{informed}, \emph{finite},
\emph{parametric}, and \emph{nonparametric} \citep{Ghahramani2013}
respectively. Note that only the last three are adaptive. Furthermore,
the agents use a probability matching strategy known as \emph{Thompson
sampling} to pick their actions. Accordingly, in each turn the agent
samples one bias for each arm from the posterior distribution over
$\Theta$ and then plays the arm with the largest bias \citep{Thompson1933}.
Fig.~\ref{fig:model_classes} compares the predictions made by the
four probabilistic models.

The nominal rewards issued by the bandits should not be confused with
the real rewards defined in the previous section. Although nominal
rewards feature in the description of the problem setup of multi-armed
bandits, finding a policy that maximizes rewards is in general intractable.
Therefore, the resulting policies will \emph{not} be optimal with
respect to those nominal rewards. However, we can use the free energy
functional to infer the real rewards optimized by the stochastic process
defined by the interactions between the constructed agents and the
bandits.

\subsection{Present Scope and Perceived Durations}

We measure the passage of time using clocks, e.g.\ hourglasses, wristwatches,
planetary movements and atomic clocks, all of which register changes
in the physical state of the world \citep{Riggs2015}. Analogously,
an agent tracks the passage of time through the changes in its memory
state triggered by the interactions. Maintaining a memory state induces
temporal correlations between the past and the future that are a signature
of the underlying adaptive mechanisms~\citep{Bialek2001a,Shalizi2001,Still2012}.
Conversely, the lack of temporal correlations is indicative of the
absence of memory.

\begin{figure}
\begin{centering}
\includegraphics[width=1\textwidth]{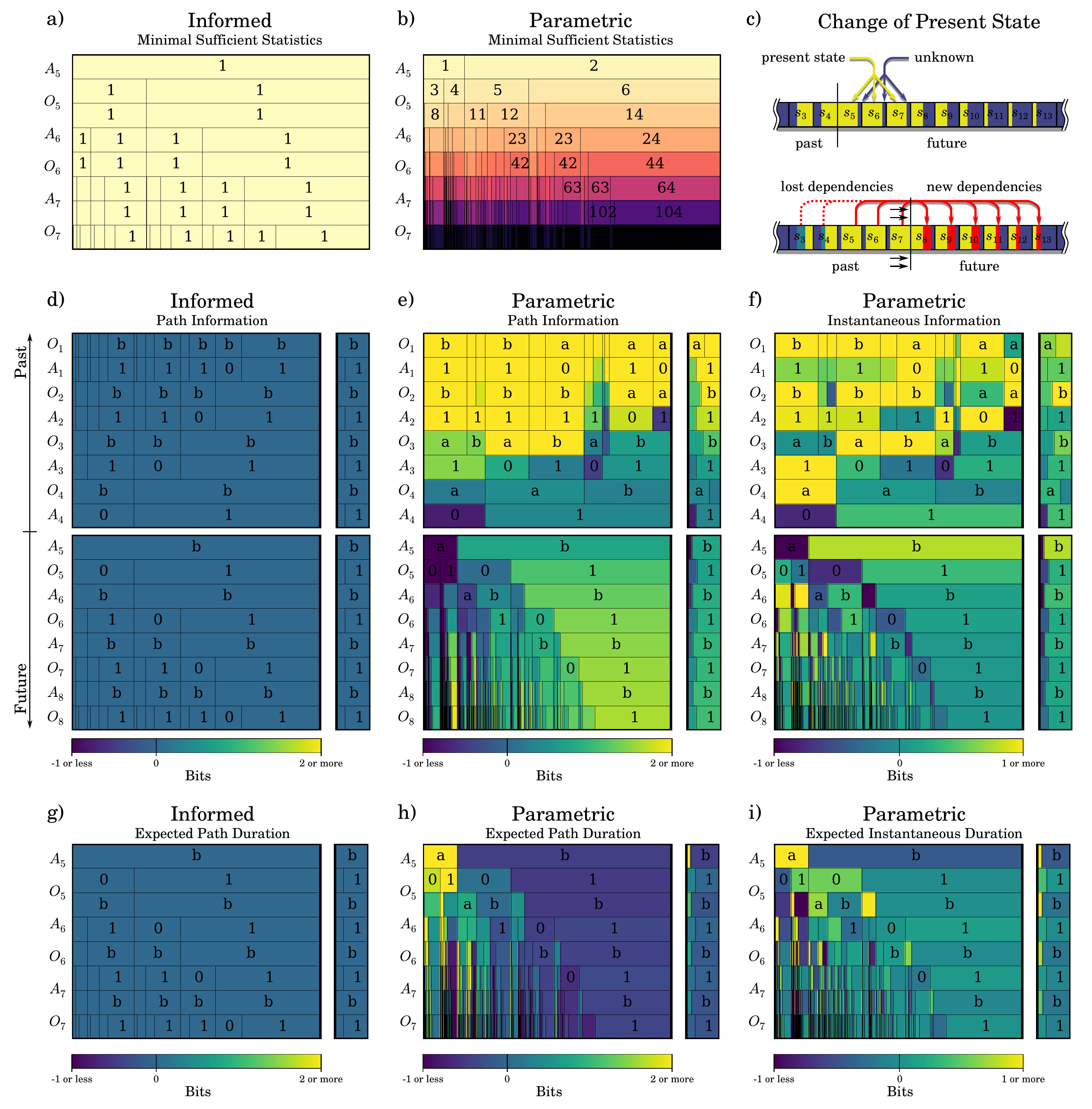}
\par\end{centering}

\caption{(a,b)~Comparison of the minimal sufficient statistic (i.e.~the memory
state) of two probabilistic models. In the diagrams, the states are
sequentially numbered and color-coded to highlight repetitions. The
probabilistic model of the informed agent has only a single memory
state, whereas the parametric class possesses many. (c)~The duration
of a sequence of interactions is defined by taking the difference
in information between two different memory states. The experience
of new interactions causes the agent to acquire new predictive information
about the future while forgetting details about the past. (d--f)~Illustration
of the agent's present scope measured in terms of the information
shared between the memory state and the past/future paths. In~(d)
\&~(e), the cell color encodes the cumulative information for the
path starting from the past-future divide and then leading up to the
cell. Diagram~(f) is obtained from~(e) by taking the difference
between two consecutive paths. (g,h)~Expected duration of interaction
paths. The cell color encodes the total duration of the path taken.
(i)~Expected duration of individual interactions, obtained as the
difference between the duration of two consecutive paths. In all diagrams,
we conditioned the models on the past $x_{\text{p}}=(\text{a1,a0,b1,b1})$
as in Fig.~\ref{fig:model_classes}. \label{fig:time-perception}}

\end{figure}

At any given point during the realization of the stochastic process,
it is natural to define the ``present'' as the minimal information
contained in the memory state that can be confirmed empirically, i.e.~the
minimal sufficient statistic of the past (Fig.~\ref{fig:time-perception}a,b).
Imagine that the agent experiences the past $x_{\text{p}}$ and enters
state $s$. In this state, the amount of information the agent possesses
about any particular future $x_{\text{f}}$ is equal to 
\begin{equation}
\text{Present(\ensuremath{x_{\text{f}}})}=\log\frac{P(x_{\text{f}}|x_{\text{p}})}{P(x_{\text{f}})}=\log\frac{P(x_{\text{f}}|s)}{P(x_{\text{f}})}\label{eq:present}
\end{equation}
bits. Similarly, only $\log\bigl[P(x_{\text{p}}|s)/P(x_{\text{p}})\bigr]$
bits about the past are remembered. Both of these quantities are upper
bounded by $-\log P(x_{\text{p}})$, the total information contained
in the past. The scope of the agent's present is the part of the past/future
that is remembered/predicted by the agent's memory state (Fig.~\ref{fig:time-perception}d--f).
In the special case of the informed agent, the present state has zero
span (Fig.~\ref{fig:time-perception}d) because it does not need
to maintain any memory in order to behave optimally. In contrast,
the parametric agent possesses an extensive scope (Fig.~\ref{fig:time-perception}e,f)
that grows with more experience. In particular, the agent can only
remember its past up to a permutation of the interactions because
the distribution over observations is exchangeable. Furthermore, it
predicts futures that are consistent with the past experience; for
instance, in the illustrated case the agent's most likely future repeatedly
pulls arm ``b'' and observes ``1'' in accordance with the past $x_{\text{p}}=(\text{a1,a0,b1,b1})$.
Deviations from the past, such as those caused by oddballs, contradict
the memory state (i.e.~they share a negative number of bits). Note
that the possession of a present scope is a property shared by all
adaptive agents.

Given this definition of the present scope, we propose to model the
passage of time relative to the stochastic process as the number of
bits in the memory that change during the experience of an interaction
(Fig.~\ref{fig:time-perception}c). This definition rests upon the
assumption that the stochastic process is implemented on a computation
model having fixed bandwidth per operation---such as a (probabilistic)
Turing machine, which can only modify a limited number of bits per
cycle%
\footnote{In the case of a probabilistic Turing machine, the only bits that
can change within a cycle are those needed to represent the state
of the Turing machine, the movement the header, and the content the
current cell in the tape.%
} \citep{Sipser1996,Papadimitriou1994}. We argue that this provides
a more plausible time metric than the time index or even the entropy
of the stochastic process, for using the index of the stochastic process
would yield a non-homogeneous complexity per time unit%
\footnote{Since an individual interaction can be arbitrarily complex, forcing
it to be computed in one time unit would require a machine that can
operate at an unbounded speed.%
}. Formally, given a finite past $x_{\text{p}}$ and future $x_{\text{n}}x_{\text{f}}$,
consider a transition that increases the past from $x_{\text{p}}$
to $x_{\text{p}}x_{\text{n}}$ and reduces the future from $x_{\text{n}}x_{\text{f}}$
to $x_{\text{f}}$ as depicted in Fig.~\ref{fig:time-perception}c.
The perceived duration of $x_{\text{n}}$ relative to this limited
window is equal to the difference 
\begin{align}
\text{Duration(\ensuremath{x_{\text{n}}})} & =\log\frac{P(x_{\text{f}}|x_{\text{n}},x_{\text{p}})}{P(x_{\text{f}})}-\log\frac{P(x_{\text{f}},x_{\text{n}}|x_{\text{p}})}{P(x_{\text{f}},x_{\text{n}})}\nonumber \\
 & =\log\frac{P(x_{\text{n}}|x_{\text{f}})}{P(x_{\text{n}}|x_{\text{p}})}.\label{eq:duration}
\end{align}
These durations are illustrated in Fig.~\ref{fig:time-perception}g--i.
The informed agent operates in the equilibrium regime and thus does
not experience time (Fig.~\ref{fig:time-perception}g). In contrast,
the parametric agent perceives durations that vary with the current
knowledge. For instance, Fig.~\ref{fig:time-perception}h--i show
that predictable interactions decrease in duration (e.g.~``b1,b1,b1''),
whereas a mistake (e.g.~accidentally playing ``a'' instead of ``b'')
and some of rewards result in longer durations. Oddballs, such as
the deviations from ``b1,b1,b1'', do \emph{not} necessarily entail
longer durations.

\subsection{Temporal Discounting}

When an agent has a limited capacity to predict the future, it can
only exploit a fraction of the rewards that lie ahead. The precise
amount can be inferred by inspecting how much they affect the behavior
of the stochastic process. If the probability $P(x_{\text{f}}|x_{\text{p}})$
of choosing a future $x_{\text{f}}$ given a past $x_{\text{p}}$
is the result of optimizing the free energy functional \eqref{eq:free-energy},
then the change in log-probability can be written as (Materials \&
Methods)

\begin{equation}
\log\frac{P(x_{\text{f}}|x_{\text{p}})}{P(x_{\text{f}})}=\beta\bigl[R(x_{\text{f}}|x_{\text{p}})+F(x_{\text{p}},x_{\text{f}})-F(x_{\text{p}})\bigr].
\end{equation}
Using~\eqref{eq:present}, we identify the l.h.s.~with the information
about the future $x_{\text{f}}$ predicted by the present. The r.h.s.~is
proportional to the difference between two terms: $R(x_{\text{f}}|x_{\text{p}})+F(x_{\text{p}},x_{\text{f}})$,
the cumulative and terminal rewards of $x_{\text{f}}$; and $F(x_{\text{p}})$,
the certainty-equivalent value of all the potential futures. This
result states that the change of a choice's probability depends exclusively
on how much it improves upon the summarized value of the choice set.
In accordance to decision theory, we refer to this excess as the \emph{rejoice}
(i.e.~negative regret, see Materials \& Methods) \citep{Bleichrodt2015}.
The rejoice scales proportionally with the agent's memory and is therefore
determined by the learning model. 

In the economic literature, an agent that reacts only to a fraction
of the reward is explained through a discount function. Typically,
a discount function gives smaller weights to rewards that lie farther
in the future \citep{Frederick2002}. Following the same rationale,
we next show how to derive discount functions that re-weight the rewards
in order to equate them with the rejoices. 

\begin{figure}
\begin{centering}
\includegraphics[width=1\textwidth]{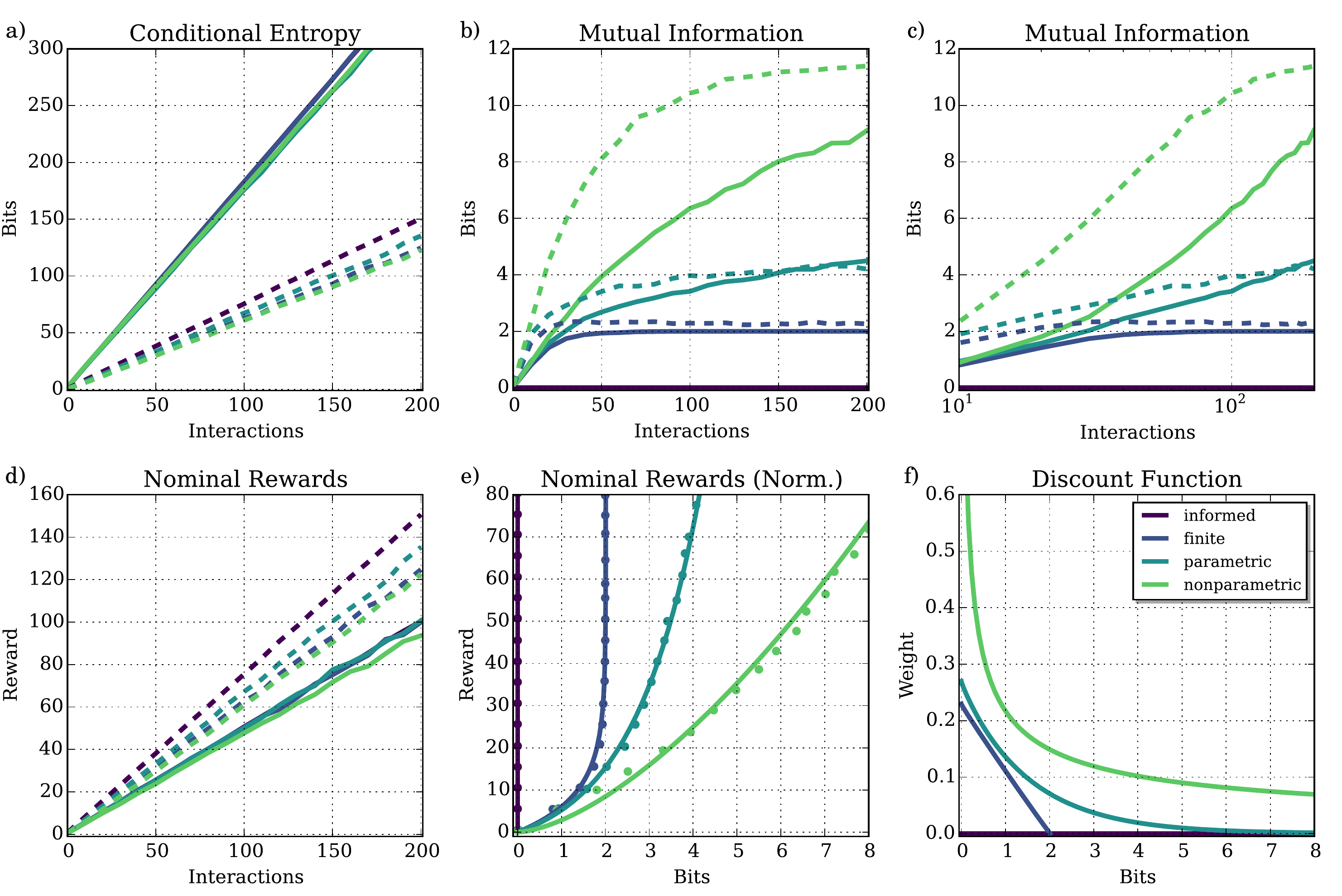}
\par\end{centering}

\caption{Monte-Carlo averages of agent-environment systems. The color-codes
are the same throughout the plots. Agents used either a random policy
(solid curves) or Thompson sampling (dashed curves). (a) Conditional
entropy of the past given the future $H(X_{\text{p}}|X_{\text{f}})$.
This curve represents the average over the negative, real rewards
of equation \eqref{eq:likelihood-rewards}. (b,c) Mutual information
between the past and future. The log-scale in (c) was chosen so as
to emphasize the logarithmic growth of the parametric case. (d)~Expected
nominal rewards of the future as a function of the length of the past-future
window. (e)~Averaged nominal rewards as a function of the mutual
information, where the latter is interpreted as a measure of perceived
duration. The plot shows the empirical data together with the fitted
curves for agents using a uniformly random policy. (f)~Discount functions
derived from the curves in~(e). They provide lower bounds for the
discount functions used by the agents using an adaptive policy. \label{fig:monte-carlo}}

\end{figure}

We conducted Monte-Carlo simulations of the four agents to study the
evolution of their predictive performance (Materials \& Methods).
To isolate the effects of the adaptive policy from the effects of
learning the environment, we also ran each agent a second time but
using a (non-adaptive) uniform random policy. All estimates were obtained
from averaging simulated trajectories of even length that were split
in the middle to get a past $x_{\text{p}}$ and a future $x_{\text{f}}$.
The results are shown in Fig.~\ref{fig:monte-carlo}. The first panel
(Fig.~\ref{fig:monte-carlo}a) shows that the conditional entropies
of the past given the future $H(X_{\text{p}}|X_{\text{f}})$, which
were obtained by averaging over the negative log-likelihoods $-\log P(x_{\text{p}}|x_{\text{f}})$,
grow linearly with the size of the window. As discussed in Sec.~\ref{sub:fe},
log-likelihoods are affine transformations of the real rewards. In
contrast, the mutual information between the past and the future grows
sublinearly, and the growth rate increases with model complexity (Fig.~\ref{fig:monte-carlo}b--c).
The mutual information curves for non-adaptive policies provide a
lower bound for their adaptive counterparts. However, adaptive policies
converge to the optimal policies in the limit, and so we expect their
mutual information curves to converge to the lower bound. The simulations
confirm this in the finite and parametric agents. Fig.~\ref{fig:monte-carlo}c
also shows that the parametric model has a logarithmic growth in mutual
information. In comparison, the nonparametric model is seen to be
super-logarithmic and the finite model upper-bounded. Finally, we
have also inspected the sum of the nominal rewards contained in the
futures. Fig.~\ref{fig:monte-carlo}d shows that all the agents predict
nominal rewards that grow proportionally with the length of the past-future
window. The slopes of these curves vary, and agents with adaptive
policies predict better rewards.

\begin{table}
\begin{centering}
\caption{Least-Square Fits of PR and MI. \label{tab:fits}}

\par\end{centering}

\centering{}%
\begin{tabular}{lcccc}
\toprule 
 & Model $f(z)$ & $a$ & $b$ & MSE\tabularnewline
\midrule
NR, Finite & $bz$ & --- & 0.5013 & 0.3421\tabularnewline
NR, Parametric & $bz$ & --- & 0.5001 & 0.5309\tabularnewline
NR, Nonparametric & $bz$ & --- & 0.4721 & 0.0893\tabularnewline
MI, Finite & $a(1-\exp(-bz))$ & 2.0107 & 0.0567 & 0.0018\tabularnewline
MI, Parametric & $a\log(1+bz)$ & 1.5283 & 0.0884 & 0.0036\tabularnewline
MI, Nonparametric & $az^{b}$ & 0.3291 & 0.6327 & 0.0779\tabularnewline
\bottomrule
\end{tabular}
\end{table}

We computed linear fits of the nominal reward curves (NR) of Fig.~\ref{fig:monte-carlo}d
and nonlinear fits for the mutual information (MI) curves of Fig.~\ref{fig:monte-carlo}b
for the agents with uniform policies, which will serve as lower bounds
for the ones based on adaptive policies. Specifically, the nonlinear
models have the following asymptotic behavior in the length~$z$
of the interval of past interactions: exponential decay $\mathcal{O}(e^{-bz})$
for the finite agent; logarithmic growth $\mathcal{O}(\log bz)$ for
the parametric agent; and power-law growth $\mathcal{O}(z^{b})$,
$0<b<1$, for the nonparametric agent. The last two have been justified
analytically in previous work: \citet{Bialek2001a} have shown that
the predictive information of models based on finite- and infinite-dimensional
parameter vectors has logarithmic and power-law growth respectively.
Notice that since nominal and real rewards are both asymptotically
linear, choosing one or the other for our fits would lead to the same
asymptotic conclusions. The results of this fit are listed Table~\ref{tab:fits}.
Combining these fits, we calculated the predicted rewards relative
to the agent's perceived duration (Fig.~\ref{fig:monte-carlo}e).
The plots show that agents predict a superlinear growth relative to
their perceived duration. Given that the rejoice is proportional to
the duration, agents must assign decaying weights to the predicted
rewards, and these weights are given by the discount functions listed
in Table~\ref{tab:discount}, shown in Fig.~\ref{fig:monte-carlo}f
(Materials \& Methods). The discount functions can be classified according
to their asymptotic decay, resulting in: infinite, linear, exponential,
and hyperbolic discounting for the informed, finite, parametric, and
nonparametric agent respectively. These functions provide lower bounds
for the discount functions of the agents using an adaptive policy.

\begin{table}
\caption{Discount Functions\label{tab:discount}}

\begin{centering}
\begin{tabular}{lccc}
\toprule 
 & $\delta(\tau)$ & Growth & Type\tabularnewline
\midrule
Informed & $0\vphantom{\Bigl(\Bigr)}$ & 0 & Infinite\tabularnewline
Finite & $\max(0,0.2274\lyxmathsym{\textminus}0.1131\tau)\vphantom{\Bigl(\Bigr)}$ & $\mathcal{O}(1)$ & Linear\tabularnewline
Parametric & $0.2701e^{\text{\textminus}0.6543\tau}\vphantom{\Bigl(\Bigr)}$ & $\mathcal{O}(e^{-c\tau})$ & Exponential\tabularnewline
Nonparametric & $0.2314\tau^{-0.5805}\vphantom{\Bigl(\Bigr)}$ & $\mathcal{O}(\tau^{-c})$ & Hyperbolic\tabularnewline
\bottomrule
\end{tabular}
\par\end{centering}

\end{table}

\section{Discussion}

Our theoretical model relates the change of an agent's memory state
to the perceived duration of interactions and to the rejoice. Concisely
stated, this relationship is given by 

\begin{equation}
\Delta\text{Memory}=\text{Duration}=\Delta\text{Rejoice.}
\end{equation}
Specifically, we have taken the memory to be synonymous with the minimal
sufficient statistics (i.e.~encoding all past-future dependencies
and only those) of the agent's probabilistic model. As suggested by
the agent-environment setup, we expect the memory substrate in animals
to encompass every modulator of behavior, ranging from the synaptic
weights in the brain to the immediate surroundings. Furthermore, an
increase of the information-processing capacity per interaction can
be related to physiological factors such a decrease in body size or
an increase in metabolic rate \citep{Healy2013}. Our agent-environment
simulations illustrate that while complex reactive behavior can emerge
from limited to no memory (as in the informed agent), adaptive behavior
depends on the availability of sufficient memory resources. 

Consistent with previous findings in the field of time perception
\citep{Eagleman2009}, the model links the shortening durations of
repeated stimuli to their increased predictability. More specifically
however, it distinguishes between three cases. Interactions that are
(a) well-predicted or (b) novel but irrelevant for future behavior
imply fewer changes to the agent's memory when experienced. Therefore,
their perceived duration is shorter. In contrast, (c) oddballs that
are \emph{relevant} for future behavior (i.e.~eliciting larger rejoice)
induce adaptation and are thus perceived as being longer in duration.
This connection between the perception of time and rewards is supported
by empirical findings in the literature on attention \citep{Brown2010}.
For instance, in a recent study where participants performed a prospective
timing task, it was found that only oddballs signaling relatively
high reward compared to the standards were perceived to last longer,
whereas oddballs with no or little reward remained unaffected \citep{Failing2016}. 

Within this context, it is instructive to examine two limit cases.
Consider an agent with a clock that ticks once per interaction. If
the agent has little to no memory (like the informed agent in our
experiments), then the temporal resolution vanishes and the clock
spins infinitely fast from the agent's point of view. In contrast,
if the agent possesses no capacity constraints, as is assumed in the
perfect rationality paradigm~\citep{Rubinstein1998}, then the clock
slows down to a point where it appears frozen. In this sense, memory
capacity acts a form of temporal inertia that quantifies the amount
of information required to move the agent one clock tick in time.

The memory constraints limit the agent's ability to react to distant
rewards, giving rise to the phenomenon of temporal discounting. The
simulation results have shown how model complexity qualitatively affects
the asymptotic behavior of discount rates; in particular, exponential
and hyperbolic discounting were shown to arise from parametric and
nonparametric model classes respectively. Given that humans have been
shown to display hyperbolic discounting \citep{Laibson1997}, our
results suggest that human intertemporal value judgment may arise
from a memory formation rate comparable to that of nonparametric models. 

The model of time perception can also account for some effects of
general memory manipulation. For instance, an increase of memory plasticity
will correlate positively with perceived durations. This phenomenon
is consistent with experimental findings, e.g.~in which the administration
of dopamine has led to the overestimation of durations and the attenuation
of impulsivity (steep discounting) \citep{Kayser2012,Joutsa2015}.

Finally, is worth remarking that the relation between memory changes,
duration and rejoice that we have laid out here is not specific to
said variables, but rather a general property of quantities that are
extensive/additive in the interactions. In other words, the limitations
imposed by the memory growth rate appear to be a general property
of perception. Verifying this property for other perceptual modalities
(other than time and reward) is a task to be further explored in the
future.

\section*{Materials and Methods}

\paragraph{Multi-Armed Bandit Processes.\label{par:mab}}

We considered simple agents and environments where the set of actions
and observations were chosen as $\mathcal{A}=\{\text{a},\text{b}\}$
and $\mathcal{O}=\{0,1\}$ respectively. For the actions, the symbols
``a'' and ``b'' encode the left and the right arm respectively; for
the observations, 0 and 1 correspond to a nominal loss and reward
responses. The environment is a two-armed bandit characterized by
a bias vector $\theta=[\theta_{\text{a}},\theta_{\text{b}}]^{T}$,
where $\theta_{a}\in[0,1]$ for $a\in\{\text{a, b}\}$. When the agent
pulls arm $a$, the bandit replies with a reward drawn from a Bernoulli
distribution with bias $\theta_{a}$. When the biases are known, as
is the case in the \emph{informed} agent, then the optimal strategy
consists in always playing the arm with the highest bias, $a^{\ast}=\arg\max_{a}\{\theta_{a}\}$.
In our simulations of the informed agent, we chose a bias vector equal
to $\theta=[\frac{1}{4},\frac{3}{4}].$ Hence, the optimal strategy
was to pick $a=\text{b}$ in every turn. 

When the biases are unknown, the agent's uncertainty is modeled by
placing a prior distribution over the bias vector \citep{Duff2002,Hutter2004}.
In the case of the \emph{finite} agent, the hypothesis class is given
by the set of four bias vectors $\Theta=\{\frac{1}{4},\frac{3}{4}\}\times\{\frac{1}{4},\frac{3}{4}\}$,
and it places the prior pmf
\[
f(z;\alpha,\beta):=\frac{z^{\alpha-1}(1-z)^{\beta-1}}{\sum_{\xi=\frac{1}{4},\frac{3}{4}}\xi^{\alpha-1}(1-\xi)^{\beta-1}}
\]
over each arm's bias. The terms $\alpha>0$ and $\beta>0$ are hyperparameters
that keep track of the total number of times that the bandit responded
with rewards and a losses respectively \citep{OrtegaBraun2010c,Chapelle2011}.
The resulting prior pmf is a product distribution 
\[
P(\theta)=f(\theta_{\text{a}};\alpha_{\text{a}},\beta_{\text{a}})\, f(\theta_{\text{b}};\alpha_{\text{b}},\beta_{\text{b}}).
\]
The \emph{parametric} agent enriches the finite case by extending
the hypothesis class to all the bias vectors in the unit square $\Theta=[0,1]\times[0,1]$,
placing an independent Beta pdf
\[
\mathcal{B}(z;\alpha,\beta):=\frac{z^{\alpha-1}(1-z)^{\beta-1}}{\int_{0}^{1}\xi^{\alpha-1}(1-\xi)^{\beta-1}\, d\xi}
\]
over each vector component. Because both agents have priors that are
conjugate to the Bernoulli distribution, their posteriors are obtained
by just updating the four hyperparameters $\alpha_{\text{a}}$, $\beta_{\text{a}}$,
$\alpha_{\text{a}}$, and $\beta_{\text{b}}$. For example, if the
agent plays arm ``a'' and the environment replies with a reward, then
the hyperparameter $\alpha_{\text{a}}$ is incremented in one count
and the others are kept equal. 

The \emph{nonparametric} agent is based on a more flexible model.
Rather than assuming a single bandit, the model considers a sequence
of two-armed bandits labeled as $N=1,2,3,\ldots$, each one having
its own vector of biases $[\theta_{\text{a}}(N),\theta_{\text{b}}(N)]$.
Starting at bandit $N=1$, the agent moves to the next bandit $N\rightarrow N+1$
whenever it receives a reward, and returns to the first bandit ($N=1$)
when it receives a loss. In each interaction, only the hyperparameters
of the current bandit are updated. The resulting hypothesis class
is given by the set $\Theta$ of all maps $\theta:\mathbf{N}\rightarrow[0,1]\times[0,1]$.

To generate actions, all the agents employ either a uniform strategy
or Thompson sampling~\citep{Thompson1933}. In the latter case, an
agent generates Monte-Carlo samples of the biases of the current bandit
from its posterior distribution. Subsequently, it plays the arm associated
to the largest bias.

At the beginning of each simulation, all the hyperparameters of an
agent's probabilistic model where initialized to one, i.e.~$\alpha=1$,
$\beta=1$ for the finite and parametric agents, and $\alpha(N)=1$,
$\beta(N)=1$ for all $N\in\mathbb{N}$ in the case of the nonparametric
agent. This corresponds to a uniform distribution over every bias.
Furthermore, the true (unknown) biases were sampled uniformly in accordance
to the agents' priors.

\paragraph{Sufficient Statistics.}

The sufficient statistics for the four agents follow directly from
their probabilistic models. The informed agent's stochastic process
is memoryless; therefore, its sufficient statistic can be modeled
as a constant function of the past. The predictions made by the finite,
parametric, and nonparametric agents depend upon the total counts
of rewards and losses (i.e.~the hyperparameters) observed so far,
the number $N$ of the bandit, and the action issued during the current
interaction if available.

\paragraph{Rejoice.}

The main difference between expected utility theory \citep{Neumann1944,Savage1954}
and regret theory \citep{Fishburn1982,Bell1982,Loomes1982} is that
in the former decision-makers maximize the expected utility, whereas
in the latter decision-makers minimize \emph{regret}, i.e.~they choose
an action $a$ that minimizes a function $\mathcal{Q}\{U(a),U(a_{\text{ref}})\}$,
where $a_{\text{ref}}$ is a reference action and $U$ is a utility
function. The regret function quantifies how much the utility $U(a)$
of $a$ is affected by what would have happened had $a_{\text{ref}}$
been chosen instead of $a$ \citep{Bleichrodt2015}. Arguably, the
simplest regret function is given by the difference $\mathcal{Q}\{U(a),U(a_{\text{ref}})\}=U(a_{\text{ref}})-U(a)$.

Decision-making based on the free energy functional can be related
to regret theory. The solution to the average free energy functional
is given by \emph{Gibbs distribution}

\begin{equation}
P(x_{\text{f}}|x_{\text{p}})=\frac{1}{Z}P(x_{\text{f}})\exp\biggl\{\beta\Bigl[R(x_{\text{f}}|x_{\text{p}})+F(x_{\text{p}},x_{\text{f}})\Bigr]\biggr\},\label{eq:ed}
\end{equation}
where the normalizing constant $Z$ is the partition function. It
is well-known that the optimal free energy is equal to 
\[
F(x_{\text{p}}):=\max_{\tilde{P}}F(x_{\text{p}})[\tilde{P}]=\frac{1}{\beta}\log Z.
\]
In the economic literature, this quantity is known as the certainty-equivalent,
and it is a function of the set of future rewards $\{R(y_{\text{f}}|x_{\text{p}})+F(x_{\text{p}},y_{\text{f}})\}_{y_{\text{f}}}$
that measures the agent's subjective worth of the cumulative rewards
that lie in the future. The value is bounded as 
\[
\mathbb{E}\biggl[R(X_{\text{f}}|X_{\text{p}})+F(X_{\text{p}},X_{\text{f}})\biggl|x_{\text{p}}\biggr]\leq F(x_{\text{p}})\leq\max_{x_{\text{f}}}\biggl\{ R(x_{\text{f}}|x_{\text{p}})+F(x_{\text{p}},x_{\text{f}})\biggr\},
\]
where the lower and upper bounds are attained when $\beta\rightarrow0$
and $\beta\rightarrow\infty$ respectively. Rearranging \eqref{eq:ed}
as
\[
\log\frac{P(x_{\text{f}}|x_{\text{p}})}{P(x_{\text{f}})}=\beta\Bigl[R(x_{\text{f}}|x_{\text{p}})+F(x_{\text{p}},x_{\text{f}})-F(x_{\text{f}})\Bigr],
\]
reveals that the changes in choice probabilities are governed by a
rejoice (negative regret) function that contrasts the rewards of future
realizations against the certainty-equivalent:
\[
\log\frac{P(x_{\text{f}}|x_{\text{p}})}{P(x_{\text{f}})}=-\beta\mathcal{Q}\Bigl\{ R(x_{\text{f}}|x_{\text{p}})+F(x_{\text{p}},x_{\text{f}}),\, F(x_{\text{f}})\Bigr\}.
\]

\paragraph{Monte-Carlo Simulations.}

The curves for the conditional entropy (Fig.~\ref{fig:monte-carlo}a),
the mutual information (Fig.~\ref{fig:monte-carlo}b \&~c), and
the predicted rewards (Fig.~\ref{fig:monte-carlo}d) were obtained
from Monte-Carlo averages made at equally spaced locations ($t=1,11,21,\ldots,201$).
To calculate an estimate at location $t$, we averaged the log-probabilities
and rewards of interaction sequences of length~$2t$ generated from
the agent's stochastic process described in the previous section.
Given the $n$-th simulated interaction sequence, let $x_{\text{p}}^{(n)}$
and $x_{\text{f}}^{(n)}$ be its first and second half respectively.
Furthermore, let $R_{\text{nom}}(x_{\text{f}}^{(n)}|x_{\text{p}}^{(n)})$
denote the nominal rewards over the second half, calculated as the
sum of the observations in $x_{\text{f}}^{(n)}$. The entropies $H(X_{\text{p}},X_{\text{p}})$,
$H(X_{\text{f}}|X_{\text{p}})$, $H(X_{\text{f}})$, and the expected
nominal rewards $\mathbb{E}\bigl[R_{\text{nom}}(X_{\text{f}}|X_{\text{p}})\bigr]$
were estimated as 
\begin{align}
H(X_{\text{p}},X_{\text{f}}) & \approx-\frac{1}{N}\sum_{n=1}^{N}\log P(x_{\text{p}}^{(n)},x_{\text{f}}^{(n)}),\label{eq:mc-hx}\\
H(X_{\text{f}}|X_{\text{p}}) & \approx-\frac{1}{N}\sum_{n=1}^{N}\log\frac{P(x_{\text{p}}^{(n)},x_{\text{f}}^{(n)})}{P(x_{\text{p}}^{(n)})},\label{eq:mc-ce}\\
H(X_{\text{f}}) & \approx-\frac{1}{N}\sum_{n=1}^{N}\log\biggl\{\frac{1}{M}\sum_{m=1}^{M}\frac{P(x_{\text{p}}^{(m)},x_{\text{f}}^{(n)})}{P(x_{\text{p}}^{(m)})}\biggr\},\label{eq:mc-me}\\
\mathbb{E}\bigl[R_{\text{nom}}(X_{\text{f}}|X_{\text{p}})\bigr] & \approx\frac{1}{N}\sum_{n=1}^{N}R_{\text{nom}}(x_{\text{f}}|x_{\text{p}})^{(n)}.\label{eq:mc-pr}
\end{align}
We used the difference between \eqref{eq:mc-me} and \eqref{eq:mc-hx}
as an estimate for the mutual information $I(X_{\text{p}};X_{\text{f}})=H(X_{\text{f}})-H(X_{\text{f}}|X_{\text{p}})$.
In particular, note that the estimate of the marginal entropy $H(X_{\text{f}})$
was obtained using a doubly-stochastic Monte-Carlo average over $NM$
samples in which the second half was kept fixed during blocks of size
$M$. The conditional entropy of the past given the future, which
stands for a proxy of the real rewards, was estimated using the formula
$H(X_{\text{p}}|X_{\text{f}})=H(X_{\text{p}},X_{\text{f}})-H(X_{\text{f}})$.
The number of samples were chosen as $N=1000$ and $M=3000$.

In the cases of the parametric and nonparametric agents, we made additional
approximations to the Thompson sampling strategy. Specifically, we
used the normal approximation to the Beta distribution:
\[
\mathcal{B}(z;\alpha,\beta)\approx\mathcal{N}\biggl(z;\mu=\frac{\alpha}{\alpha+\beta},\sigma^{2}=\frac{\alpha\beta}{(\alpha+\beta)^{2}(1+\alpha+\beta)}\biggr).
\]
The approximation holds well for large values of $\alpha$ and $\beta$,
and it has the advantage of keeping both the generation and evaluation
of action samples computationally tractable. The probability of choosing
arm ``a'' then becomes equal to $1-F(c)$ , where $F$ is the cdf
of the normal distribution with zero mean and unit variance, and $c=\frac{1}{2}(\mu_{\text{a}}-\mu_{\text{b}})\sqrt{\sigma_{\text{a}}^{-2}+\sigma_{\text{b}}^{-2}}$.

\paragraph{Temporal discounting.}

The discount functions were derived using the following procedure.
The choice of the functional forms of the model classes listed in
Table~\ref{tab:fits} were motivated by prior studies of the long-term
behavior of the entropy and the predictive information \citep{Bialek2001a}.
The exception is the functional form of the finite model's mutual
information, which was chosen through inspection of the curve. The
parameters were fit (least-square regression) to the data obtained
in the Monte-Carlo simulations. This yielded two functions per agent:
the expected nominal reward function $R(z)$ and the mutual information
$I(z)$, both as a function of the number of interactions $z\in(0,\infty)$
of the past and the future window. For analytical convenience, we
extended the number of interactions form the discrete to the continuous
domain. The two functions $R(z)$ and $I(z)$ were then connected
via $z$. Fig.~\ref{fig:monte-carlo}e is obtained by plotting $R(\tau):=R(I^{-1}(\tau))$,
where the inverse $I^{-1}(\tau)$ is the number of interactions as
a function of the mutual information $\tau\in(0,\infty)$, now interpreted
as a temporal coordinate. These inverses were equal to: $z=\frac{1}{b}\log(\frac{a}{a-\tau})$
for the finite model; $z=\frac{1}{b}(e^{\tau/a}-1)$ for the parametric
model; and $z=(\frac{\tau}{a})^{1/b}$ for the nonparametric model.
The respective nominal reward functions $R(\tau)$ are just rescaled
versions of $I^{-1}(\tau)$. To obtain the discount functions, we
must make sure that the discounted future grows proportionally in
$\tau$, that is 
\[
\int_{t=0}^{t=\tau}\delta(t)\frac{\partial R}{\partial t}(t)\, dt=\alpha\tau,
\]
because the rejoice $\alpha\tau$ is proportional to the mutual information.
Assuming w.l.g. that $\alpha=1$, this is achieved when
\[
\delta(t)=\Bigl(\frac{\partial R}{\partial t}\Bigr)^{-1}(t).
\]
Thus, the resulting discount functions have the shapes: $\delta(\tau)=d-c\tau$
for the finite case; $\delta(\tau)=de^{-c\tau}$, for the parametric
case; and $\delta(\tau)=d\tau^{-c}$ for the nonparametric case. Note
that the constants $c$ and $d$ are positive in each case.

\subsection*{Competing Interests}

The authors declare that the research was conducted in the absence
of any commercial or financial relationships that could be construed
as a potential conflict of interest.

\subsection*{Acknowledgements}

P.A. Ortega would like to thank D.~Balduzzi, D.A.~Braun, J.R.~Donoso,
K.-E. Kim and D.~Polani for helpful comments and suggestions. This
study was funded by the Israeli Science Foundation center of excellence,
the DARPA MSEE project and the Intel Collaborative Research Institute
for Computational Intelligence (ICRI-CI).

\footnotesize

\bibliographystyle{plainnat}
\bibliography{bibliography}

\end{document}